\documentclass[%
 reprint,
 amsmath,amssymb,
 aps,
]{revtex4-2}

\usepackage{dcolumn}
\usepackage{bm}
\usepackage{hyperref}

\usepackage[dvipsnames]{xcolor}
\usepackage[T1]{fontenc} 
\usepackage{subcaption}

\usepackage[compat=1.1.0]{tikz-feynhand}
\usepackage{tikz-feynman}
\tikzfeynmanset{compat=1.1.0}
\usepackage{feynmp}
\usepackage{tikzsymbols}
\usepackage{array}
\usepackage{pifont} 
\usepackage{soul}
\usepackage{float}
\usepackage{comment}
\usepackage[utf8]{inputenc}
\usepackage{rotate}
\usepackage{graphicx}
\usepackage{amsmath}
\usepackage{amssymb}
\usepackage[normalem]{ulem}
\usepackage{multirow}
\usepackage{mathrsfs} 
\usepackage{adjustbox}
\usepackage{mathtools}

\begin{document}

\preprint{2108.xxxx}

\title{\boldmath \color{BrickRed} Paths to proton stability in grand unification}

\author{Lisa L. Everett}
 \email{leverett@wisc.edu}
 \affiliation{Department of Physics, University of Wisconsin-Madison, Madison, WI 53706}
\author{Oleg Popov}%
 \email{opopo001@ucr.edu}
\affiliation{%
 Department of Physics, Korea Advanced Institute of Science and Technology, \\291 Daehak-ro, Yuseong-gu, Daejeon 34141, Republic of Korea
}%
\affiliation{
 Scuola Normale Superiore, Piazza dei Cavalieri 7, 56126 Pisa, Italy
}%


\date{\today}

\begin{abstract}
We explore scenarios within grand unification in which proton decay can be suppressed and possibly eliminated due to novel embeddings of the Standard Model matter fields into irreducible representations of the grand unified group and residual symmetries. The scenarios are based on an $SU(7)$ gauge group, in which the matter fields are embedded within an anomaly-free set of fields that can be realized as a natural subgroup of the fundamental spinor representation of an $SO(14)$ gauge symmetry.  Depending on the embedding, proton decay can either be forbidden at tree level and generated via one-loop diagrams, or the proton can be made stable by forcing it to decay channels that must have an even number of leptons, independently of the bosonic content of the theory. We describe the theoretical and phenomenological implications of such scenarios, including their implications for dark matter and neutrino masses.  
%
\end{abstract}

\maketitle


%
%
{\it Introduction}. Grand unified theories (GUTs) provide an elegant unification of the gauge and matter fields of the Standard Model (SM), as in the famous models based on $SU(5)$ \cite{Georgi:1974sy}, $SO(10)$~\cite{Georgi:1974my,Fritzsch:1974nn}, and $E(6)$ gauge groups~\cite{Gursey:1975ki} (see~\cite{Langacker:1980js} for an early review; an updated review can be found in~\cite{ParticleDataGroup:2020ssz}). However, in models with economical embeddings of the SM matter fields into irreducible representations of the grand unified group, there is inevitable tree-level proton decay mediated by vector bosons, which in turn requires high-scale GUT symmetry breaking. The limit on the proton lifetime is currently at the level of $>10^{34}$ yrs~\cite{ParticleDataGroup:2020ssz}, placing severe constraints on minimal implementations of the grand unification paradigm.

The challenge of rendering the proton stable or sufficiently stable in grand unification has long been known (see e.g.~\cite{Gell-Mann:1976xin,Langacker:1980js}).  
One option is to consider direct product groups rather than a simple grand unified group, such as the trinification~\cite{Rujula:1984abc,Babu:1985gi} and quartification~\cite{Babu:2003nw,Joshi:1991yn,Kownacki:2017kuq,Kownacki:2017uyq} models, as well as the original Pati-Salam model~\cite{Pati:1974yy}. In simple grand unified groups, proton decay can be controlled via conserved combinations of gauge and global symmetries that distinguish quarks from leptons \footnote{For example, in~\cite{Langacker:1977ai,Langacker:1978fn}, the conserved quantum number that distinguishes quarks from leptons is the ``chromality'' associated with global $U(1)_c$.}. Another approach is to consider nonminimal embeddings of the SM quarks and leptons into GUT multiplets, so that tree-level vector-mediated diagrams are absent. This idea was recently explored within $SU(5)$ in~\cite{Fornal:2017xcj}, where the chiral $\mathbf{\Bar{5}}$ and $\mathbf{10}$ are augmented by vectorlike pairs of fermions in higher-dimensional $SU(5)$ representations.

In this paper, we explore paths within grand unification in which proton decay channels are controlled by novel embeddings of the SM quarks and leptons into GUT multiplets and residual symmetries. We work for concreteness within $SU(7)$ grand unification, with a specific anomaly-free set of $SU(7)$ irreducible representations. 
Depending on the embeddings and the symmetry breaking patterns, the proton can either decay at a suppressed rate to specific two-body channels (either $B+L$ conserving or $B-L$ conserving), or it can decay only to three-body channels, or it can be made stable. We show two limiting scenarios that realize these features and discuss their theoretical and phenomenological implications. 
%
%

{\it Minimal field content}. We take $SU(7)$ gauge symmetry as the grand unified group, with the minimal fermion and scalar field content as given in Table~\ref{tab:fields_su7main}. 
\begin{table}[H]
    \centering
    \begin{tabular}{ccccccccc}
        \hline \hline
        $SU(7)$ &  & ${\mathbf{\Bar{7}}}_F$ & $\mathbf{21}_F$ & ${\mathbf{\Bar{35}}}_F$ &  & $\mathbf{7}_S$ & $\mathbf{48}_S$ & $\mathbf{84}_S\; \mathbf{210}_S$ \\ 
        \hline \hline
    \end{tabular}
    \caption{The minimal field content and its $SU(7)$ representations for the fermion ($F$) and scalar ($S$) fields.}
    \label{tab:fields_su7main}
\end{table}
We assume that $SU(7)$ is broken to the SM gauge group in several steps. The first step is to break $SU(7)$ via the vacuum expectation value (VEV) of an adjoint scalar field, $\mathbf{48}_S$, as follows:

\begin{equation}
    SU(7)\rightarrow SU(3)_c\times SU(2)_L \times \left[U(1)\right]^3,
    \end{equation}
    in which $U(1)^3$ is  $U(1)_F\times U(1)_{1^\prime}\times U(1)_Y$, where $U(1)_Y$ is the usual SM hypercharge. The two additional $U(1)$ groups are sequentially broken as follows: $U(1)_{1^\prime}$ is first broken by the vacuum expectation value of a SM singlet component of $\mathbf{7}_S$, and $U(1)_F$ is then broken either by an $\mathbf{84}_{S}$ or $\mathbf{210}_{S}$. The breaking of the SM gauge group to $SU(3)_c\times U(1)_{\rm em}$ occurs via the electroweak Higgs field, which is in a $\mathbf{7}_S$ representation of $SU(7)$. 
%

We further assume that the SM quarks and leptons of a single generation are embedded in the anomaly-free set of $SU(7)$ irreducible representations (irreps) given by $\mathbf{\Bar{7}}_F$, $\mathbf{21}_F$, and $\mathbf{\Bar{35}}_F$. The $\mathbf{\Bar{7}}^j$ is the antifundamental; its decomposition in terms of $SU(3)_c\times SU(2)_L\times U(1)^3$ is given by
$\pmb{\Bar{7}} \sim \left(\pmb{\Bar{3}}, \pmb{1}, -\frac{1}{3}, -1, \frac{1}{3}\right) \oplus \left(\pmb{1}, \pmb{1}, \frac{5}{3}, -1, 0\right) \oplus \left(\pmb{1}, \pmb{1}, 0, 6, 0\right) \oplus \left(\pmb{1}, \pmb{2}, -\frac{1}{3}, -1, -\frac{1}{2}\right)$. 
$\mathbf{21}_{ij}$ is the rank-two antisymmetric tensor, and $\mathbf{\Bar{35}}^{ijk}$ is the conjugate of the rank-three antisymmetric tensor (their decompositions are easily obtained using the result for the fundamental irrep). For the scalars, $\mathbf{48}^i_j$ is the adjoint, $\mathbf{7}_{i}$ is the fundamental, and $\mathbf{84}_{ijk}$ and $\mathbf{210}_{ijkl}$ are symmetric tensors.  We employ $SU(7)$ indices $\{ 1,1^\prime, 2,3\}$, with lower (upper) indices labeling standard (conjugate) irreps. 
The Yukawa interactions of the fermions and the $\mathbf{7}_S$ are
\begin{align}
\label{eq:lagrangian}
    -\mathscr{L}^{\text{Yuk}} &= Y_d \Bar{7}_F 21_F 7_S^{\dagger} + Y_u 21_F \Bar{35}_F 7_S \nonumber \\ &+ Y_e \Bar{35}_F \Bar{35}_F 7_S^{\dagger}+ \text{h.c.}
\end{align}
There are $63$ fermions, which in analogy to $SU(5)\rightarrow SO(10)$, can be extended by a $SU(7)$ singlet fermion to $64=2^{7-1}$ and interpreted as the components of the fundamental spinor $\mathbf{64}_F$ of $SO(14)$, of which $SU(7)$ is a natural subgroup \cite{Georgi:1979md,Umemura:1980sa}. 
The fermions are divided into two sets, each consisting of two copies of SM-like fermions and their mirrors, which have conjugate SM charges. 
For each set, the $U(1)_F$ charges are also conjugate, but the $U(1)_{1'}$ charges are chiral. This chiral $U(1)$ reflects the original chirality of this combination of $SU(7)$ representations. The two additional $U(1)$ gauge groups can be identified with the $U(1)$'s that arise upon the breaking of $SU(7)$ first to $SU(6)\times U(1)_6$, then $SU(6)$ breaks to $SU(5)\times U(1)_5$, and the SM gauge group arises as usual from the breaking of $SU(5)$ \footnote{See \cite{Umemura:1980sa} for an alternative breaking chain and hypercharge embedding.}.  Here the identification is that $U(1)_{F}=U(1)_5$, and $U(1)_{1'}=U(1)_6$. The $SU(7)$ fermion representations can be broken down into $SU(5)$ labels as $\mathbf{\Bar{7}}_F = \mathbf{1}_F\oplus \mathbf{1}_F\oplus \mathbf{\Bar{5}}_F$, $\mathbf{21}_F = \mathbf{1}_F\oplus \mathbf{5}_F\oplus \mathbf{5}_F\oplus \mathbf{10}_F$, and $\mathbf{\Bar{35}}_F=\mathbf{\Bar{5}}_F\oplus \mathbf{\Bar{10}}_F\oplus \mathbf{10}_F\oplus \mathbf{\Bar{10}}_F$. The decompositions of the two copies of the SM-like states with respect to $SU(3)_c\times SU(2)_L\times U(1)^3$ are shown in Table~\ref{tab:quark_lepton_assignments} (the SM mirror states are not shown). We employ standard notation in which the SM multiplets that are $SU(2)_L$ singlets are denoted by barred quantities $\Bar{u}$, $\Bar{d}$, etc., where the bar is part of the name of the field, rather than a conjugation.
\begin{table}[H]
    \centering
    \begin{tabular}{|c|lc|cl|c|}
        \hline
        \multicolumn{6}{|c|}{$SU(3)_c \times SU(2)_L \times U(1)_F \times U(1)_{1^\prime} \times U(1)_Y$} \\ \hline
        $Q_1$ & $\pmb{21}_{32}$ & $\left(\pmb{3}, \pmb{2}, \frac{2}{3}, 2, \frac{1}{6}\right)$ & $\left(\pmb{3}, \pmb{2}, -1, -3, \frac{1}{6}\right)$ & $\pmb{\Bar{35}}^{332}$ & $Q_2$ \\
        $\Bar{u}_1$ & $\pmb{\Bar{35}}^{322}$ & $\left(\pmb{\Bar{3}}, \pmb{1}, -1, -3, -\frac{2}{3}\right)$ & $\left(\pmb{\Bar{3}}, \pmb{1}, \frac{2}{3}, 2, -\frac{2}{3}\right)$ & $\pmb{21}_{33}$ & $\Bar{u}_2$ \\
        $\Bar{d}_1$ & $\pmb{\Bar{7}}^{3}$ & $\left(\pmb{\Bar{3}}, \pmb{1}, -\frac{1}{3}, -1, \frac{1}{3}\right)$ & $\left(\pmb{\Bar{3}}, \pmb{1}, \frac{4}{3}, 4, \frac{1}{3}\right)$ & $\pmb{\Bar{35}}^{311^\prime}$ & $\Bar{d}_2$ \\ \hline
        $L_1$ & $\pmb{\Bar{35}}^{211^\prime}$ & $\left(\pmb{1}, \pmb{2}, \frac{4}{3}, 4, -\frac{1}{2}\right)$ & $\left(\pmb{1}, \pmb{2}, -\frac{1}{3}, -1, -\frac{1}{2}\right)$ & $\pmb{\Bar{7}}^{2}$ & $L_2$ \\
        $\Bar{e}_1$ & $\pmb{\Bar{35}}^{333}$ & $\left(\pmb{1}, \pmb{1}, -1, -3, 1\right)$ & $\left(\pmb{1}, \pmb{1}, \frac{2}{3}, 2, 1\right)$ & $\pmb{21}_{22}$ & $\Bar{e}_2$ \\
        $\Bar{\nu}_1$ & $\pmb{21}_{11^\prime}$ & $\left(\pmb{1}, \pmb{1}, -\frac{5}{3}, -5, 0\right)$ & $\times$ & $\times$ & $\Bar{\nu}_2$ \\
        \hline
        \multicolumn{6}{|c|}{$H\sim \mathbf{7}_2 \sim \left(\pmb{1}, \pmb{2}, \frac{1}{3}, 1, \frac{1}{2}\right)$} \\
        \multicolumn{6}{|c|}{$\phi^\prime\sim \mathbf{7}_{1^\prime} \sim \left(\pmb{1}, \pmb{1}, 0, -6, 0\right)$} \\
        \hline
    \end{tabular}
    \caption{The two copies of SM fields (with one right-handed neutrino) and two of the relevant scalar field components in the $SU(7)$ scenarios considered.
    }
    \label{tab:quark_lepton_assignments}
\end{table}
%
%

A key requirement is that one of these two sets of fermions must acquire vectorlike masses of the order of the $U(1)_{1^\prime}$ breaking scale, while the other must be prevented from acquiring such large vectorlike masses, as it is to contain the SM matter content corresponding to one generation of quarks and leptons. As stated, this latter set not only includes the SM fields, but also mirror fermions, which are subject to stringent phenomenological constraints.  We will comment more shortly on the general challenge of decoupling the exotic fermions while keeping the SM fields light, which will likely require an extension of the fermion content beyond this minimal set of $SU(7)$ irreducible representations.

{\it Proton stability--general considerations}. We now consider various possibilities for how the SM quarks and leptons are embedded in the set of $SU(7)$ fermion representations as given in Table~\ref{tab:quark_lepton_assignments}. Quite generally, what states are to be identified as SM fermions and which are exotics from the perspective of the low energy effective theory will depend in detail on how the masses are generated that split the heavy vectorlike set from the remaining particles. However, in certain limiting cases, there are significant implications for the proton decay constraints.

For example, one possibility is to proceed along standard (minimal) lines, given that this breaking could be understood as a breaking that preserves an $SU(5)$ subgroup that itself breaks to the SM gauge symmetry. This standard option would be to embed the SM quarks and leptons in the $\mathbf{\Bar{5}}_F$ and $\mathbf{10}_F$ that arise from the $\mathbf{\Bar{7}}_F$ and $\mathbf{21}_F$ of $SU(7)$, respectively. In the notation of Table~\ref{tab:quark_lepton_assignments}, this  corresponds to the assignments $\Bar{d}_1$, $L_2$, $Q_1$, $\Bar{u}_2$, $\Bar{e}_2$, $\Bar{\nu}_2$. However, this choice would result in to severe proton decay constraints from unavoidable tree-level dimension six operators, just as in the case for minimal $SU(5)$. 

We thus consider alternative possibilities for the embedding of the SM quarks and leptons, in which the tree-level vector-mediated proton decay operators are absent or controlled. We focus for concreteness on the following two limiting scenarios, again in the notation of Table~\ref{tab:quark_lepton_assignments}:
\begin{itemize}
    \item {\bf Case 1}: $Q_1$, $\Bar{u}_1$, $\Bar{d}_1$, $L_1$, $\Bar{e}_1$, $\Bar{\nu}_1$.
    In this case, the $\Bar{d}$ and $L$ are in different $SU(7)$ irreps, as are the $Q$ and $\Bar{e}$ fields (in terms of the effective $SU(5)$ subgroup as described above, $\Bar{d}$ and $L$ are in different $\mathbf{\Bar{5}}$'s, and $Q$ is in a different $\mathbf{10}$ from $\Bar{u}$ and $\Bar{e}$).
    \item {\bf Case 2}: $Q_2$, $\Bar{u}_2$, $\Bar{d}_2$, $L_1$,   $\Bar{e}_1$, $\Bar{\nu}_1$. In this case, the $\Bar{d}$ and $L$ fields are in the same $SU(7)$ irrep, as are the $Q$ and $\Bar{e}$, but $u$ is in a different $SU(7)$ irrep (in $SU(5)$ subgroup language, $\Bar{d}$ and $L$ are in the same $\mathbf{\Bar{5}}$, and $Q$ and $\Bar{e}$ are in the same $\mathbf{10}$).
\end{itemize}
For both cases, tree-level vector-mediated proton decay is avoided, as the diquark interactions are absent at tree level, though leptoquark interactions are present (the lepton embeddings are the same). This can be understood either in terms of the $SU(5)$ subgroup as described above, or by noticing that the multiple $SU(7)$ index changes required among the relevant matter fields for these interactions cannot occur via the adjoint vector field, as the adjoint can only raise or lower one index at a time. 

However, the two cases vary widely in terms of their implications for proton stability.  In Case 1, we will see that proton decay occurs at one-loop, with decay channels dictated by the selection rules from the $U(1)_F$ symmetry and its relation to the effective baryon number $B$ and lepton number $L$ global symmetries, and the scalar fields that break the $U(1)_F$ symmetry.  Once the $U(1)_{1'}$ is broken and the exotics are decoupled, an effective $Z_2$ symmetry emerges, with the $Z_2$-odd exotic states propagating in the loops.  For Case 2, in contrast, we will see that a selection rule arises that renders the proton stable, due to a conflict between the requirements of the $U(1)_F$ symmetry and its connection to $B$ and $L$, and the requirements of Lorentz invariance.


%

%
%
%
{\it Proton stability: Case 1}. 
With the assignment of SM quarks and leptons for Case 1, tree-level proton decay diagrams are not generated either via vector gauge boson interactions or via any tree-level interactions with scalar fields, as long as the SM fields do not mix with the mirror states, as required. However, one-loop diagrams arise in which the heavy states with masses set by the breaking of the $U(1)_{1'}$ symmetry propagate in the loops. These processes can be classified into the following two general categories: (i) decays which preserve $B-L$, and are proportional to the vacuum expectation value of the field ($\mathbf{84}_S$ or $\mathbf{210}_S$) that breaks the $U(1)_F$ symmetry, with corresponding implications for the decay channels, and (ii) decays which preserve $B+L$ and, while independent of the vacuum expectation value of the $U(1)_F$-breaking field, require the inclusion of additional scalar fields that are even rank partially or fully symmetric tensor irreducible representations of $SU(7)$, such as $\mathbf{28}_S$ or $\mathbf{196}_S$.

These features can be understood by considering how the $U(1)_F$ gauge symmetry can be interpreted as a specific combination of the effective $B$ and $L$ symmetries and an additional global accidental symmetry, given by $U(1)_X$, and examining the effective $SU(2)_G$ global symmetry of the Yukawa Lagrangian of Eq.~(\ref{eq:lagrangian}) in the low energy limit. In Table~\ref{tab:global_sym_case_1}, we show the decomposition of the SM quarks and leptons in terms of these symmetries. 
\begin{table}[H]
    \centering
    \begin{tabular}{|ccccc|c|}
        \hline
        Field & $SU(2)_G$ & $U(1)_B$ & $U(1)_L$ & $U(1)_X$ & $U(1)_F$ \\ \hline
        $\left(\begin{matrix} Q_1 & L_1\end{matrix}\right)^T$ & $\pmb{2}$ & $\left(\begin{matrix} \frac{1}{3} & 0\end{matrix}\right)^T$ & $\left(\begin{matrix} 0 & 1\end{matrix}\right)^T$ & $\frac{1}{3}$ & $\left(\begin{matrix} \frac{2}{3} & \frac{4}{3}\end{matrix}\right)^T$ \\
        $\left(\begin{matrix} \Bar{u}_1 & \Bar{\nu}_1\end{matrix}\right)^T$ & $\pmb{2}$ & $\left(\begin{matrix} -\frac{1}{3} & 0\end{matrix}\right)^T$ & $\left(\begin{matrix} 0 & -1\end{matrix}\right)^T$ & $-\frac{2}{3}$ & $\left(\begin{matrix} -1 & -\frac{5}{3}\end{matrix}\right)^T$ \\
        $\left(\begin{matrix} \Bar{d}_1 & \Bar{e}_1\end{matrix}\right)^T$ & $\pmb{2}$ & $\left(\begin{matrix} -\frac{1}{3} & 0\end{matrix}\right)^T$ & $\left(\begin{matrix} 0 & -1\end{matrix}\right)^T$ & $0$ & $\left(\begin{matrix} -\frac{1}{3} & -1\end{matrix}\right)^T$ \\
        \hline
        $H$ & $\pmb{1}$ & $0$ & $0$ & $\frac{1}{3}$ & $\frac{1}{3}$ \\
        \hline
    \end{tabular}
    \caption{Accidental global symmetries for Case 1.}
    \label{tab:global_sym_case_1}
\end{table}
In this case, we can identify the quantum number $F$ of $U(1)_F$ as
\begin{align}
    \label{eq:u1f_x_rel}
    F &= B + L + X. 
\end{align}
In the infrared limit, Eq.~(\ref{eq:lagrangian}) becomes
\begin{eqnarray}
    -\mathscr{L}_{\text{IR}} &=& Y_{u\nu} \left(\begin{matrix}Q_1 & L_1\end{matrix}\right)\left(\begin{matrix}\Bar{u}_1 \\ \Bar{\nu}_1\end{matrix}\right) H \\&+& Y_{de} \left(\begin{matrix}Q_1 & L_1\end{matrix}\right)\left(\begin{matrix}\Bar{d}_1 \\ \Bar{e}_1\end{matrix}\right) H^\dagger + \text{h.c.}, \nonumber
\end{eqnarray}
leading to a scenario similar to the one observed in Pati-Salam models ($SU(4)_c \times SU(2)_L \times U(1)_X$). 

Let us now consider the effective proton decay operators with all gauge and global symmetries conserved.
Examples of such operators include
\begin{eqnarray}
    \label{eq:p_decay_ope_1}
    \left(\Bar{35}^{\dagger}_{211^\prime} \Bar{\sigma}^\mu 21_{32}\right) \left(\Bar{7}^{\dagger}_{3}\Bar{7}^{\dagger}_{3}\right)
    &\sim& 
    \left(L_1^\dagger \Bar{\sigma}^\mu Q_1\right) \left(\Bar{d}^\dagger_1 \Bar{d}^\dagger_1\right), 
    \nonumber \\
    \left(\Bar{35}^\dagger_{322} \Bar{\sigma}^\mu 21_{11^\prime}\right)\left(\Bar{7}^{\dagger}_{3}\Bar{7}^{\dagger}_{3}\right) 
    &\sim& 
    \left(\Bar{u}^\dagger_1 \Bar{\sigma}^\mu \Bar{\nu}\right)\left(\Bar{d}^\dagger_1 \Bar{d}^\dagger_1\right),
    \\
    \left(\Bar{7}^\dagger_3 \Bar{\sigma}^\mu \Bar{35}^{333}\right)\left(\Bar{7}^{\dagger}_{3}\Bar{7}^{\dagger}_{3}\right)
    &\sim& 
    \left(\Bar{d}^\dagger_1 \Bar{\sigma}^\mu \Bar{e}_1\right)\left(\Bar{d}^\dagger_1 \Bar{d}^\dagger_1\right). 
    \nonumber
\end{eqnarray}
These effective operators contribute to proton decay channels that preserve $B+L$, such as
    $p \rightarrow \nu \pi^+$,
    $p\rightarrow e^- \pi^+ \pi^+$, etc. 
From Table~\ref{tab:global_sym_case_1} and Eq.~(\ref{eq:u1f_x_rel}), we see that the operators of Eq.~(\ref{eq:p_decay_ope_1}) 
conserve both the $U(1)_F$ gauge symmetry and the $U(1)_X$ global symmetry, and thus preserve $B+L$. However,  these operators violate the $SU(2)_G$ global symmetry, and as a result are forbidden. To generate these operators, what is needed is to introduce additional scalar fields that couple to fermions in such a way that violates $SU(2)_G$ explicitly and softly. For example, adding a scalar field in the symmetric rank-two tensor representation $\mathbf{28}_S$ leads  to the following key couplings: 
\begin{equation}
    \label{eq:coupling_28}
    \Bar{7}_F 28_S \Bar{7}_F + 7_S 28^{\dagger}_S 7_S 
    \sim 
    \Bar{d}_1 28_{S,33} \Bar{d}_1 + 7_{S,3} 28^{\dagger 33}_S 7_{S, 3}. 
\end{equation}
These interactions violate $SU(2)_G$ and thus induce $B+L$ conserving proton decays in both two-body and three-body body decay channels, such as $p\rightarrow v \pi^+$, $p \rightarrow e^+ \nu \nu$, $p \rightarrow e^- \pi^+ \pi^+$. Similar arguments apply to other operators that induce $B+L$ conserving proton decays via the presence of such new scalar fields. 

Hence, all proton decay channels in Case 1 can be divided into two categories. The first are $B+L$ conserving, which are all proportional to the couplings involving the new scalar field(s), as in Eq.~(\ref{eq:coupling_28}). The second are $B-L$ conserving processes, which require $U(1)_{F}$ violation. The proton decay width via $B-L$ conserving channels is proportional to the the vacuum expectation value of the scalar field that breaks the $U(1)_F$ gauge symmetry, such as the $\langle \mathbf{84}_{111'}\rangle$. A variety of one-loop topologies for the proton decay diagrams are possible, and the detailed constraints will also depend on the scale of $U(1)_{1'}$ breaking, which sets the mass scale for the $Z_2$-odd exotic states propagating in the loop. The details of this scenario, including constraints from gauge coupling unification, will be presented elsewhere \cite{Everett:2021def}.


{\it Proton stability: Case 2.} 
In this case, tree-level proton decay diagrams are absent as discussed above, but the stronger statement can be made that loop-level proton decay diagrams are also not generated.  This arises from the relation between $B$, $L$, and $F$ charges for this case, which takes the form
\begin{align}
    \label{eq:FBL_rel_2}
    F=B-{\color{BrickRed} \frac{{\color{black}L}}{2}}+\gamma X.
\end{align}
The charges of the SM quarks and leptons, as well as the SM Higgs and the $\mathbf{84}_{111^\prime}$, are given in Table~\ref{tab:global_sym_case_2}. 
\begin{table}[H]
    \centering
    \begin{tabular}{|cccc|c|}
        \hline
        Field & $U(1)_B$ & $U(1)_L$ & $U(1)_X$ & $U(1)_F$ \\ \hline
        $Q_2 $ & $\frac{1}{3}$ & $0$ & $-\frac{4}{3\gamma}$ & $-1$ \\
        $\bar{u}_2 $ & $-\frac{1}{3}$ & $0$ & $\frac{1}{\gamma}$ & $\frac{2}{3}$ \\
        $\bar{d}_2 $ & $-\frac{1}{3}$ & $0$ & $\frac{5}{3\gamma}$ & $\frac{4}{3}$ \\
        $L_1 $ & $0$ & $1$ & $\frac{11}{6\gamma}$ & $\frac{4}{3}$ \\
        $\bar{e}_1 $ & $0$ & $-1$ & $-\frac{3}{2\gamma}$ & $-1$ \\
        $\bar{\nu}_1 $ & $0$ & $-1$ & $-\frac{13}{6\gamma}$ & $-\frac{5}{3}$ \\
        \hline
        $H$ & $0$ & $0$ & $\frac{1}{3\gamma}$ & $\frac{1}{3}$ \\
        $\left\langle 84_{111^\prime}\right\rangle$ & $0$ & $0$ & $-\frac{10}{\gamma 3}$ & $-\frac{10}{3}$ \\
        \hline
    \end{tabular}
    \caption{Accidental global symmetries for Case 2.}
    \label{tab:global_sym_case_2}
\end{table}
To see this, consider an effective operator $\mathcal{O}$ that consists of products of spinors and the VEVs of scalar fields, as follows:
\begin{align}
    \mathcal{O} &\sim \overbrace{\underbrace{\left[F_1 F_2 \dots\right]}_{F_{\rm spinor}=B-L/2+\displaystyle\sum_f X_f} \quad \underbrace{\left[\left\langle S_1 \right\rangle \left\langle S_2 \right\rangle \dots \right]}_{-F_{\rm spinor}=\displaystyle\sum_{S} X_{S}}}^{\displaystyle\sum_i F=0}.
\end{align}
As $\sum F=0$ due to gauge invariance, we then must have $\displaystyle\sum_S F_S = -\displaystyle\sum_{\rm spinor} F_{\rm spinor}$. We will further assume that all scalar fields have $B=L=0$ (\emph{i.e.}, no Majorons). 

Let us first suppose for simplicitly that all scalars have integer $F$ charges, such that the sum of the spinor $F$ charges must also be an integer.  As the $\gamma$ in the $X$ charge listings in Table~\ref{tab:global_sym_case_2} is arbitrary, we can choose it so that the $X$ charges of all the involved fields are integers as well (e.g.~$\gamma=1/6$). We then see from Eq.~(\ref{eq:FBL_rel_2}) that
\begin{equation}
\underset{\in \mathbb{Z}}{F_{\rm spinor}} = B-\frac{L}{2} + \underset{\in \mathbb{Z}}{X_{\rm spinor}}\rightarrow \mathbb{Z}=B-\frac{L}{2}.   
\end{equation}
For proton decay, $B=1$, and thus we arrive at
\begin{equation}
\mathbb{Z}=\frac{L}{2}\rightarrow 2\times\mathbb{Z}=L,
\end{equation}
so that in order for proton to decay, there must be an even number of leptons in the final state. However, this contradicts Lorentz symmetry conservation, which requires an odd number of leptons in the final state. As there are no even $L$ charged leptons in the SM, we conclude that the proton will be stable if $\displaystyle\sum_S F_S \in \mathbb{Z}$.
The analogous conclusion is easily proven for the case at hand in which the scalars have fractional $F$ charges, such as $F_H = 1/3$ as in Table~\ref{tab:global_sym_case_2}. The sum of the spinor $F$ charges will then also be $\in \mathbb{Z}/3$, and we can as before choose $\gamma$ such that the $X$ charges of the involved fields are also fractions of $3$ (e.g.~$\gamma=1/2$). We now have
\begin{equation}
  \underset{\in \mathbb{Z}/3}{F_{\rm spinor}} = B-\frac{L}{2} + \underset{\in \mathbb{Z}/3}{X_{\rm spinor}}\rightarrow \frac{\mathbb{Z}}{3}=B-\frac{L}{2}.
\end{equation}
This yields
\begin{equation}
\mathbb{Z}=3-\frac{3L}{2}\rightarrow\mathbb{Z}=\frac{3L}{2}\rightarrow 2\times\mathbb{Z}=3L,    
\end{equation}
requiring an even number of leptons in the final state, in conflict with Lorentz invariance. Examples of forbidden operators due to these conflicting constraints are
\begin{align}
    \label{eq:ex1_scen_2}
    &(21_{33} 21_{33}) (\bar{35}^{311^\prime} \bar{35}^{333}) (21_{11^\prime} \times) 
    \sim (\bar{u}_2 \bar{u}_2) (\bar{d}_2 \bar{e}_1) (\bar{\nu}_1 \times), \nonumber \\
    &(\bar{35}^{332} \bar{35}^{332}) (\bar{35}^\dagger_{311^\prime} \bar{35}^\dagger_{333}) (21^{\dagger 11^\prime} \times) 7_2 7_2 \nonumber \\ &\sim (Q_2 Q_2) (\bar{d}_2^\dagger \bar{e}_1^\dagger) (\bar{\nu}_1^\dagger \times) H H,  \\
    &(\bar{35}^{332} \bar{35}^{332}) (\bar{35}^\dagger_{311^\prime} \bar{35}^\dagger_{333}) (\bar{35}^{211^\prime} \times) 7_2 7_2 7_2 \nonumber \\ &\sim (Q_2 Q_2) (\bar{d}_2^\dagger \bar{e}_1^\dagger) (L_1 \times) H H H, \nonumber
\end{align}
where the $\times$ indicates a missing state required by Lorentz invariance.

Let us now imagine that the theory includes a scalar with a half-integer $F$ charge ($F_S \in \mathbb{Z}/2$). We would have
\begin{equation}
    \frac{\mathbb{Z}}{2} = B - \frac{L}{2} + \frac{X}{2} 
    = B - \frac{L}{2} \rightarrow 
    \mathbb{Z} = 2 - L.
\end{equation}
This relation implies that proton decay to any odd number of leptons would be possible if there were one or more half$-$integer $F$ charged scalars with non-vanishing VEV(s). However, 
we see from Table~\ref{tab:global_sym_case_2} that all scalar fields in the fundamental irrep of $SU(7)$ in this scenario have $F$ charges that are fractions of $3$, which implies that the $F$ charges of all higher rank irreps will be also fractions of $3$ (including the cases with integer charges). 
There are thus no scalar fields with the quantum numbers needed to reconcile the constraints of gauge and Lorentz symmetry for the generation of proton decay operators in Case 2. 
It is this property and the factor of $1/2$ in Eq.~(\ref{eq:FBL_rel_2}) that makes the proton stable in Case 2, independently of the bosonic field content of the theory.

%
%

{\it Discussion and Conclusions}. 
The extent to which either of these limiting cases can be realized depends on the details of how the fermions acquire their masses. The chiral $U(1)_{1^\prime}$ protects the states from Georgi's survival hypothesis \cite{Georgi:1979md}, \emph{i.e.}, from acquiring ultraheavy masses upon the breaking of $SU(7)$ to  $SU(3)_c\times SU(2)_L\times U(1)^3$. However, once $U(1)_{1'}$ is broken via the VEV of $\mathbf{7}_{S,1'}$, such masses, now proportional to the  $\mathbf{7}_{S,1'}$ VEV via the Yukawa interactions of Eq.~(\ref{eq:lagrangian}), are allowed. As stated above, a key requirement is that one set of fermion vector pairs acquires such masses, while the second set of fermions, which includes the SM chiral matter and a set of mirror fermions with this minimal set of $SU(7)$ irreps, must be prevented from doing so. The mirror states, if they persist to experimentally accessible energies, must not be allowed to mix with the SM fermion at any discernible level to be consistent with electroweak precision data (see e.g.~\cite{Langacker:1988ur,Langacker:1991zr,Maalampi:1988va,ParticleDataGroup:2020ssz})). If the masses of the mirror fermions are controlled by chiral couplings to the SM Higgs, they are subject to direct collider search bounds (that generally depend on the mixing of these states with the SM fermions) and perturbativity bounds. Such chiral states are also severely constrained by LHC Higgs data, for the same reasons that a fourth sequential generation of matter is ruled out in the SM \cite{Djouadi:2012ae} and strongly constrained in theories with extended electroweak Higgs sectors, such as two Higgs doublet models \cite{Das:2017mnu,Kang:2018jem}. 

One option is to extend the fermion content by including degrees of freedom that can pair up with the mirror states. For example, adding the chiral anomaly-free set $3(\mathbf{\Bar{7}})\oplus \mathbf{21}$ can allow for this possibility, along the lines proposed in \cite{Fornal:2017xcj} for $SU(5)$ \footnote{This set of matter fields was also considered in early literature \cite{Frampton:1979cw,Umemura:1980sa}; see also \cite{Fonseca:2015aoa}.}. This would require further scalar fields and parameter tuning to achieve a viable pattern of mass eigenstates. Another is to extend the theory such that the mirror fermions are prevented from sizable mixing with the SM states and their masses are made consistent with electroweak precision and LHC data, for which there may be small viable parameter regions for sufficiently extended electroweak Higgs sectors.  To maintain the proton decay features, any such mechanism to generate masses for the exotic particles must preserve the effective $Z_2$ symmetry that results after the breaking of the $U(1)_{1'}$ gauge symmetry. This symmetry also has interesting potential implications for dark matter, with possible candidates in either the scalar, fermion, or vector sectors. It also can be of importance for  neutrino mass generation, for example via the Dirac seesaw \cite{Ma:2016mwh}, or, in the presence of the $\mathbf{28}_S$ field, various mechanisms for Majorana masses, including 
scotogenic scenarios \cite{Ma:2006km}.  

In conclusion, we have explored paths to proton stability within grand unification which rely on nonstandard embeddings of the SM fermions and resulting residual symmetries, taking $SU(7)$ as the grand unified group. We identified two scenarios which do not exhibit tree-level proton decay and have specific consequences for proton stability. One predicts one-loop proton decay with heavy exotics in the loops due to an emergent $Z_2$ symmetry. The other predicts a stable proton due to a clash between the requirements of gauge and Lorentz invariance for proton decay processes. The viability of such scenarios is predicated on successfully decoupling the exotic fields from the SM fields within the $SU(7)$ fermion irreps, which is a nontrivial challenge for such nonminimal SM fermion embeddings. With stringent proton decay bounds perhaps suggesting that the elegant and minimal standard GUT models may require revision, we believe such approaches can provide an intriguing path forward for continued probes of the grand unification paradigm.


%
%

%
\begin{acknowledgments}
LLE is supported by the U. S. Department of Energy under the contract de-sc0017647. OP is supported by the Samsung Science and Technology Foundation under Grant No.~SSTF-BA1602-04 and National Research Foundation of Korea under Grant No.~2018R1A2B6007000.
\end{acknowledgments}

%
%
%

\bibliography{references}

\end{document}